Title: **Effect of the social environment on olfaction and social skills in WT and mouse model of autism**

Caroline GORA[1*], Ana DUDAS[1*], Lucas COURT[1], Anil ANNAMNEEDI[1,2,3], Gaëlle LEFORT[4], Thiago-Seike PICORETI-NAKAHARA[5], Nicolas AZZOPARDI[1], Adrien ACQUISTAPACE[5], Anne-Lyse LAINE[6], Anne-Charlotte TROUILLET[5], Lucile DROBECQ[1], Emmanuel PECNARD[1], Benoît PIEGU[4], Pascale CREPIEUX[1,7], Pablo CHAMERO[5] and Lucie P. PELLISSIER[1#]

[1] Team biology of GPCR Signaling systems (BIOS), CNRS, IFCE, INRAE, Université de Tours, PRC, F-37380, Nouzilly, France.

[2] LE STUDIUM Loire Valley Institute for Advanced Studies, 45000, Orléans, France

[3] current address: Department of Biotechnology, School of Bioengineering, SRM Institute of Science and Technology, Kattankulathur-603203, Tamilnadu, India

[4] CNRS, IFCE, INRAE, Université de Tours, PRC, F-37380, Nouzilly, France.

[5] Team Neuroendocrine Integration of Reproduction and Behavior (INERC), CNRS, IFCE, INRAE, University of Tours, PRC, F-37380, Nouzilly, France.

[6] Platform LPE, Unité PRC, Centre INRAE Val de Loire, Nouzilly, France

[7] Inria, Inria Saclay-Ile-de-France, Palaiseau, France

* Authors contributed equally

# Corresponding author: Lucie P. Pellissier, PhD, Team biology of GPCR Signaling systems (BIOS), CNRS, IFCE, INRAE, Université de Tours, PRC, F-37380, Nouzilly, France. Phone: +33 4 47 42 79 62. Email: lucie.pellissier@inrae.fr

running title: **Social isolation normalizes *Shank3* knockout phenotype**

**ABSTRACT**

Autism spectrum disorders are complex, polygenic and heterogenous neurodevelopmental conditions, imposing a substantial economic burden. Genetics are influenced by the environment, specifically the social experience during the critical neurodevelopmental period. Despite efficacy of early behavior interventions targeted specific behaviors in some autistic children, there is no sustainable treatment for the two core symptoms: deficits in social interaction and communication, and stereotyped or restrained behaviors or interests. In this study, we investigated the impact of the social environment on both wild-type (WT) and *Shank3* knockout (KO) mice, a mouse model that reproduces core autism-like symptoms. Our findings revealed that WT mice raised in an enriched social environment maintained social interest towards new conspecifics across multiple trials. Additionally, we observed that 2 hours or chronic social isolation induced social deficits or enhanced social interaction and olfactory neuron responses in WT animals, respectively. Notably, chronic social isolation restored both social novelty and olfactory deficits, and normalized self-grooming behavior in *Shank3* KO mice. These results novel insights for the implementation of behavioral intervention and inclusive classrooms programs for children with ASD.

**INTRODUCTION**

Autism spectrum disorder (ASD) represents a debilitating neurodevelopmental disease, carrying a substantial economic and societal burden. Diagnosis is primarily based on two core behavioral symptoms: deficits in social interaction and communication and repetitive, restrained and stereotyped behaviors or interests[1]. Often, ASD is accompanied with comorbid symptoms, such as anxiety, epilepsy, motor and cognitive deficits and sleep perturbations[1]. Over the past decade, the prevalence of ASD has reached 1% of the global population[2–4]. Genome-wide association studies have identified over a thousand susceptibility or risk genes, with the *SHANK3* gene standing out in the high confidence and syndromic category ([SFARI](SFARI)). Located within the 22q13 region deleted in the syndromic Phelan-McDermid syndrome frequently associated with ASD[5], the *SHANK3* gene has been linked to ASD through 142 reports, with more than 340 mutations or rare variants ([SFARI](SFARI)). Consequently, the *Shank3b* KO mouse serves as a valuable model for studying ASD, effectively mirroring core symptoms of the condition. The phenotype includes social novelty deficits (e.g., reduced interaction with an unknown mouse compared to a familiar one), along with exacerbated motor stereotypies[6]. With half of ASD cases remaining idiopathic, genetics are influenced by environmental factors and gene dosage, as indicated by the variability in ASD severity among twins or siblings[7]. Thus, ASD stands as a complex, polygenic, and heterogenous disorder.

So far, no pharmacological treatment or cure addresses social deficits and/or stereotyped core symptoms in individuals with ASD. Approved medications for ASD include atypical antipsychotics, such as risperidone and aripiprazole, as well as antidepressants, targeting irritability, self-injury, rituals, anxiety and depression[8–10]. However, they have limited efficacies and come with consequent side effects. Despite extensive testing of numerous compounds targeting receptors or signaling pathways[11], they have consistently failed in clinical trials. Their lack of efficacy can be attributed to the large placebo effect and the inherent diversity among patients. Treatment options hinge on the development of chemical and biochemical compounds that target G protein coupled receptors - highly tunable proteins that control dysregulated signaling pathways in ASD. This holds promise for new therapeutic treatments in subtypes of ASD patients[11].

Applied Behavior Analysis (ABA)-based interventions for ASD children have demonstrated some efficacy in addressing core social and cognitive deficits compared to standard child care services[12–14]. However, these interventions are both expensive and demanding, requiring 35 hours per week of intensive training for both children and their caregivers[12]. It is recommended that ABA interventions start as early as possible, ideally, before 18 months[13], which aligns with the challenges associated with early access to diagnosis. Moreover, these interventions improve only the targeted behavior[14]. Interestingly, the translation of ABA to *Oprm1* KO young adult mice has shown improvement, specifically in targeted social interaction[15]. Despite these findings, the outcomes of ABA interventions in ASD children remain uncertain, preventing broader implementation across all ASD children and family[16,17]. Alternatives may involve combining professionally adapted interventions with integrating ASD and neurotypical children in inclusive classrooms. While positive outcomes have been associated with neurodiversity among classroom participants and teacher engagement compared to autism-specific classrooms[18–20], the implementation and sustainability of such approaches remains challenging. Indeed, inappropriate implementation within inclusive school programs results in significant behavioral difficulties for ASD children, often leading them to discontinue those programs and join specialized schools[21,22].

During the COVID pandemic, children and adolescents have experienced social isolation, ranging from separation from their peers to complete loneliness, particularly during a crucial period for the establishment of social relationships. Reports indicate increased aggression and depression during or shortly after chronic social isolation[23], underscoring the critical impact of such unprecedented global isolation, including its implications for individuals with ASD. Studies in rodents reveal that acute (24 hours) or chronic isolation can have opposing effects on sociability, either enhancing social seeking for a conspecific or leading to social deficits and aggression, respectively (reviewed in[23]). Recent findings in rats demonstrate that acute (1 day) or short-term (7 days) social isolation rapidly induces social memory impairment[24]. Despite the restoration of behavioral deficits, dysregulation in neural plasticity genes and proteins (e.g., Bdnf, Egr1, Arc) persists after reunion[24]. These observations highlight the significant impact of the duration of social isolation on sociability. However, the effects of social isolation on ASD mouse models have yet to be explored. Therefore, this study aims to determine the impact of early social environment on WT animals and chronic social isolation on *Shank3* KO mice, a mouse model useful for the study of ASD-like core symptoms.

**MATERIALS AND METHODS**

*Animals*

All mouse breeding, care and experimental procedures were in accordance with the European and French Directives and approved by the local ethical committee CEEA Val de Loire N°19 and the French ministry of teaching, research and innovation (APAFIS #18035-2018121213436249). 2 months-old naïve *Shank3* KO (JAX stock #017688)[6] and control wildtype (WT) males and females were maintained on a mixed (50%-50%) C57BL/6J;129S2 background. Mice were either raised in groups of 2-4 individuals (gradient of social enrichment) or isolated for 2 hours (acute isolation) or for 4 weeks (chronic isolation) in a clean bedding cage prior to behavioral phenotyping. All animals were maintained in the same experimental room on a 12hr regular light/dark cycle, with food and water *ad libitum* and controlled temperature (±21°C) and humidity (±50%).

*Behavioral tests*

All behavioral tests were performed under dim light in a standardized apparatus designed to reduce anxious-like behavior. Timelines, detailed procedures and scoring are fully described in the **supplementary methods**. Briefly, social interaction was assessed during 10 minutes using the reciprocal social interaction test, the three-chambered (habituation, followed by sociability and social novelty phase) and in the Live Mouse Tracker[25]. Cognitive inflexibility and perseverative, stereotyped and repetitive behaviors were tested in Y-maze task and motor stereotypy tests. Locomotion and anxious-like behaviors were measured in the open field. Olfaction was tested using an olfactory preference test.

*Quantitative PCR*

Total RNAs from olfactory tissues (Main olfactory epithelium (MOE), vomeronasal organ (VNO) and olfactory bulb (OB)) were extracted according to the manufacturer instructions (Zymo Research Corporation kit Direct-zol RNA Microprep) and quantified using a nanodrop (Thermo Fisher Scientific, Waltham, MA). cDNAs (MOE, VNO and OB) were generated from 450 ng, 320 ng and 115 ng of total RNAs, respectively, using the SuperScript III kit (Invitrogen) according to the manufacturer protocol. All qPCRs were performed in 384 well plates using cDNA (dilution 1/50 for MOE and VNO and 1/25 dilution for OB) and primers for the corresponding genes (Table S1, supplementary methods), accordingly to the manufacturer's protocol (2X Ozyme ONEGreen Fast qPCR premix).

*RNAscope In situ hybridization*

Mice were anesthetized with a mixture of 100 mg/kg ketamine and 5 mg/kg xylazine, perfused transcardially with PBS, followed by 4% PFA. VNO tissue was dissected, postfixed overnight in 4% PFA and cryoprotected in 30% sucrose. Samples were embedded in Tissue-Tek O.C.T. compound, snap-frozen in cold isopentane and processed on a Leica CM3050S cryostat. VNOs were cut in 16 μm thick coronal sections and mounted on SuperFrost Plus glass slides (Thermo Scientific). Fluorescent *in situ* hybridization (FISH) for *Oxtr* mRNA was performed according to the RNAscope Fluorescent Multiplex V2 labeling kit (ACDBio 323110) coupled to TSA signal development using the mm-*Oxtr* probe (ACDBio 412171) are fully described in supplementary methods. Briefly, slides were incubated with HRP RNAscope reagent, washed, incubated with Biotin-conjugated Tyramide 1:50, washed, incubated with Cy2-labeled streptavidin 1:500, washed, counterstained with Hoechst 33258 and mounted with antifade fluorescent mounting medium. Fluorescent images were acquired using the laser scanning confocal microscopy (Zeiss LSM-780). Quantification was performed using the "Subcellular detection" feature of QuPath software.

*Calcium imaging*

$Ca^{2+}$ imaging of freshly dissociated VSNs was performed as previously described[26,27]. VNO epithelium was detached from the cartilage and minced in PBS at 4°C. The tissue was incubated (20 min at 37°C) in PBS supplemented with papain (0.22 U/ml; Worthington) and DNase I (10 U/ml; Thermofisher), gently extruded in DMEM (Invitrogen) supplemented with 10% FBS, and centrifuged at 100 × g (5 min). Dissociated cells were plated on coverslips previously coated with concanavalin-A type V (0.5 mg/ml, overnight at 4 °C; Sigma). Cells were used immediately for imaging after loading them with fura-2/AM (5 µM; Invitrogen) for 30 min. Cell-containing coverslips were placed in a laminar-flow chamber (Warner Instruments) and constantly perfused at 22 °C with extracellular solution Hank's balanced salt solution (HBSS, Invitrogen) supplemented with 10 mM Hepes. Cells were alternately illuminated at 340 and 380 nm, and light emitted above 510 nm was recorded using a C10600-10B Hamamatsu camera installed on an Olympus IX71 microscope. Images were acquired at 0.25 Hz and analyzed using ImageJ (NIH), including background subtraction, region of interest (ROI) detection and signal analyses. ROIs were selected manually and always included the whole cell body. Peak signals were calculated from the temporal profiles of image ratio/fluorescent values. Results are based on recordings from 5 - 7 mice for each condition and genotype. Cells were stimulated successively and in random order using a bath application with control HBSS-Hepes buffer, 1 µM oxytocin (Sigma), and urine mixed from at least three different adult male mice and diluted to 1:100. To identify and mark $Ca^{2+}$ responses to the applied stimuli, we followed these criteria: 1) a response was defined as a stimulus-dependent deviation of fluorescence ratio that exceeded twice the standard deviation of the mean of the baseline fluorescence noise, 2) cells showing a response to the control buffer were excluded from analysis, and 3) a response had to occur within 1 min after stimulus application. In time series experiments, ligand application was repeated to confirm the repeatability of a given $Ca^{2+}$ response.

*Statistical analysis*

Data and statistical analysis were performed using R software (version 4.2.2). For behavioral data, Kruskal-Wallis tests followed by Dunn's *post-hoc* tests were performed with the rstatix package [Kassambara, A. 2020. rstatix: Pipe-friendly framework for basic statistical tests. (Available from: https://CRAN.R-project.org/packageprstatix)] to identify significant differences between mouse lines. For LMT data, the complete linear robust model was estimated with the rlm function of the MASS package, ANOVA were used to select the significant variables (number of mice per cage, trial or interaction between the two), *post-hoc* tests based on estimated marginal means were implemented with the emmeans package [Lenth R (2023). emmeans: Estimated Marginal Means, aka Least-Squares Means. R package version 1.8.5] to identify differences between the factors of the significant variables (for example, differences between mouse lines). Quantitative PCR analysis was performed using the BioRad CFX Maestro 2.3 software on normalized relative expression calculated using control samples and reference targets (ΔΔCq). Comparisons between-groups were done using the two-tailed Student's t test. For calcium imaging, statistical analyses were performed using the GraphPad Prism software 10.0.2 and the Fisher's exact test. The probability of error level (alpha) was chosen to be 0.05. For RNAscope puncta quantification, the raw number of dots was analyzed using GraphPad Prism with one-way ANOVA between the groups.Raw data, mean ± standard deviation (sd) and statistics are presented in **Supplementary Tables**.

**RESULTS**

**High social enrichment improves social interaction**

Previously, reports have shown that social environment may influence social skills and aggression in mice[23]. Thus, we wondered whether low, moderate to rich social environments, by housing respectively by 2, 3 and 4 wildtype (WT) animals per cage, would have a differential effect on social interaction. Using the Live Mouse Tracker that allows automatic detection of social interaction parameters[25], we observed that lower social environment decreased social interest and exploration across subsequent trials by reducing the time spent in social approach, movement in contact and nose contacts (**Figure 1**). By contrast, high social enrichment showed beneficial effects by maintaining these social parameters and reducing the time spent isolated (**Figure 1**). None of the housing conditions impacted the time spent in get away, a parameter of social avoidance, or time in stretch attend posture or in the periphery, two indexes of anxious-like behavior (**Table S2**). These results were consistent between males and females for the housing condition, although overall, females showed improved social interaction compared to males, as they spent more time in social approach, move in contact and nose contacts with the other females (**Figure S1**, **Table S2**). In conclusion, these results provide evidence that social enrichment is beneficial for increasing sociability in grouped mice.

**Chronic social isolation improves social interactions**

Next, we reasoned that WT mice exposed to social isolation may lead to social interaction deficits, as previously suggested[23]. Thus, we analyzed the behavior of mice exposed to two hours of acute social isolation and observed reduced social interactions in these animals, revealed by reduced time in nose contacts in the reciprocal social interaction tests (**Figure S2**,

**Table S3**). Surprisingly, mice subjected to 4-week chronic social isolation did not display any social interaction deficits. On the contrary, chronic isolation showed beneficial effects on social interaction by increasing the time and mean duration of nose contacts and preference for a mouse over a toy within the 10 minutes duration of those tests, an index of prosocial behavior (**Figure 2A-C**, **Table S3**). Furthermore, chronic social isolation improved nose contacts in two independent social tests, independently of sex or background of the mice despite some variability between cohorts (**Figure S3**, **Table S3**).

Chronic social isolation did not elicit any detrimental effects on other ASD symptoms, such as motor stereotypies, perseverative behaviors, cognitive flexibility or anxious-like behaviors, as the time spent in self-grooming and digging, the number of headshakes or the spontaneous alternation pattern or in the center of the open field, respectively, were unaffected (**Figures 2D-E**, **S4**, **Table S3**). However, social isolation enhanced locomotor activity as observed in the traveled distance, mean speed and the number of rearing (**Figure 2F**, **Table S3**). Consistent with previous reports[23], social isolation also induced a mild increase in male aggression, 2 out of 12 isolated males attacked only upon two consecutive exposure to male interactors in the social novelty phase of the 3-chambered tests. We also observed an effect of the mouse background, since mice from a mixed background generally displayed enhanced social skills and cognitive flexibility, and reduced motor stereotypies (**Figures S3-S4**). To conclude, chronic but not acute social isolation had a beneficial effect on social skills related to core symptoms of ASD.

**Chronic social isolation alters expression of oxytocin family genes only in the VNO**

One of the key features of chronic social isolation is the limited access to social odors from other conspecifics. Given that olfaction is one of the primary senses in mice that detect social signals, we tested whether chronic social isolation would impact olfactory function. We hypothesized that the oxytocin (OT) family genes that express in olfactory tissues[28,29] might be impacted. Indeed, social isolation induced a 14-36-fold increase in *Oxt*, *Oxtr* and *Avp* mRNA levels in the vomeronasal organ (VNO), but not in the main olfactory epithelium (MOE) or in the olfactory bulb (OB; **Figure 3**). *In situ* hybridization using an *Oxtr* probe confirmed that chronic isolation led to an enrichment of *Oxtr* mRNA levels in basal and apical layers of the VNO sensory epithelium containing vomeronasal sensory neurons (VSN), as well as in other non-sensory cells. The VNO is specialized in the detection of pheromones and other chemosignals that trigger various social behaviors, including aggression, sexual and parental behaviors, which are also modulated by oxytocin[30]. Thus, we hypothesized that mice could detect oxytocin (OT) and vasopressin (AVP) peptides and that chronic isolation would enhance those responses. Using an olfactory preference test, we observed in both isolated and grouped housing conditions a preference for OT (**Figure 4A**, **Table S4**), indicating that mice are able to detect OT. Although OT did not induce a significant difference in this cohort, *Shank3* KO mice did not show major alteration in urine preference or delay in the latency of the first olfactory contact (**Figure 4B-C**, **Table S4**), indicating similar performance than WT mice. Interestingly, WT mice raised in groups showed a strong preference for urine from isolated sex-matched animals, as an index of interest for an unknown odor (**Figure 4D-G**, **Table S4**). However, our results did not show any difference in OT concentrations in urine of isolated animals, suggesting that the olfactory preference is mediated by other chemosignals. In contrast, OT plasma concentrations were higher in isolated animals and, when available, did not correlate with urine levels. In conclusion, mice are able to detect OT in an olfactory preference test and chronic social isolation led to upregulation of oxytocin family genes specifically in the VNO.

**Chronic social isolation normalizes social deficits and stereotyped behaviors in *Shank3* KO mouse model of ASD**

*Shank3b* KO displayed social novelty impairment and exacerbated motor stereotypies (**Figure 5**, **Table S5**). We hypothesized that chronic isolation of *Shank3* KO mice could impact their social phenotype. Unexpectedly, social isolation enhanced social interaction in *Shank3* KO mice, by increasing the time spent in nose contacts with unknown animals in different social paradigms and huddling (**Figures 5A-E**, **S5A-B**, **Table S5**), an indicator of social comfort and bonds[31]. Furthermore, social isolation also normalizes some motor stereotypies in *Shank3* KO mice, such as the time spent, number and mean duration of self-grooming and digging, but not headshakes episodes. Remarkably, we observed that soon after one *Shank3* KO mouse started to perform self-grooming, other KO mice in separated and contiguous cages started to groom as well. These synchronized self-grooming existed, to a much lower extent in WT and was nearly absent in isolated WT and *Shank3* KO mice (**Figures 5F**, **S5C-D**, **Table S5**), suggesting that the exacerbated self-grooming phenotype observed in *Shank3* KO mice may be consequence of life in social groups. To our knowledge, this is the first report of synchronized self-grooming induced by social housing in a mouse model of ASD.

Next, we next wondered whether VSN responses were affected in *Shank3* KO mice and whether social isolation would also normalize those deficits. To better characterize the effect of chronic isolation in the VNO of WT and *Shank3* KO males, we analyzed VSN responses to urine and oxytocin using live-cell calcium imaging. We found that VSNs of *Shank3* KO mice responded less to urine (from 3 to 1.6%; 1.9-fold) and oxytocin (from 3 to 1.9%; 1.6-fold; **Figure 5G-I**), indicating dysfunction in this primary olfactory epithelium. Chronic isolation increased the number of VSNs responding to urine (from 3 to 5.1%; 1.7-fold) in WT mice and *Shank3* KO mice, although the increase was more pronounced (5-fold more VSNs responding to urine) but not to OT (from 3 to 3.7% and 1.9 to 2.4%, respectively).

In conclusion, chronic social isolation in *Shank3* KO mice had beneficial effects on the two core symptoms of ASD, accompanied with normalization of olfactory responses. Although it remains to be investigated, restoration of social olfactory processing and motivation in *Shank3* KO may play a role in this effect. To our knowledge, this is the first report of the effect of OT on VSN signaling, however, it is not clear how enhanced oxytocin family gene expression could impact olfaction.

**DISCUSSION**

In this study, we demonstrated that WT male and female mice exposed to acute (two hours) or chronic (4 weeks) social isolation exhibited changes in sociability (showing reduction and increases, respectively), but not in social novelty. These findings contradict previous reports[23,24]. In the case of acute isolation, animals isolated for two hours might still experience stress, potentially countering the isolation-induced social motivation and seeking for a social partner, unlike the findings from 24-hours isolation in earlier reports[23,24], explaining our results. This disparity could be linked to corticotropin-releasing hormone levels[23], especially in the paraventricular nucleus of the hypothalamus where oxytocin neurons are located. However, the divergence in chronic isolation cannot be attributed to stress or mouse background, as observed in this study, but is more likely associated with housing conditions. Chabout and colleagues demonstrated that isolated animals emit comparable amounts of ultrasound vocalizations as those raised in groups, and even more upon reunion[32]. On one hand, constant communication between animals exposed to chronic isolation and those raised in groups may have prevented the detrimental effects of chronic isolation, potentially also explaining the enhanced social motivation and seeking for conspecifics. On the other hand, well-being of animals has improved over the last decades, with successive generations raised in an enriched (social) environment, potentially mitigating the adverse effects of this stressor. Indeed, a semi-naturalistic environment improved core ASD-like symptoms in BTBR inbred mice[33]. Therefore, exploring the impact of social isolation across generations in a less enriched environment might provide clarity on this question.

Notably, our findings reveal that C57BL/6J;129S2 mixed background exhibits enhanced social interaction skills compared to the pure C57BL/6J background. This aligns with previous research. Various inbred backgrounds manifest distinct social and repetitive behaviors[34]. Moreover, recent reports indicate that hybrid or mixed backgrounds have improved behavioral skills, as observed for the sensorimotor performance in hybrid FVB/NJ;C57BL/6J background or social interaction in *Cdh8*[+/-] mouse model of ASD[35,36]. Using diverse genetic backgrounds is likely to enhance the reliability of results, particularly in the context of social interaction, thereby enhancing the characterization of mouse models of autism.

Our main discovery lies in the amelioration of both core symptoms in *Shank3* KO mice, specifically social novelty deficits and self-grooming, when exposed to social isolation. Consistent with the original study[6], we observed a substantial deficit in social novelty in *Shank3* KO mice using the 3-chambered test, accompanied with motor stereotypies. Additionally, Maloney and colleagues reported diminished social motivation in *Shank3* KO mice[37]. Therefore, chronic isolation likely reshapes brain circuitry and restores social motivation in *Shank3* KO mice, notably intensifying the seeking of a conspecific and the need for social comfort, as evidenced by the time spent huddling. Furthermore, we found that chronic isolation normalized self-grooming but not head shake episodes in *Shank3* KO mice. While the latter likely reflects motor impairment in the striatum of this model[6], self-grooming seems to be acquired through life in social groups. Specifically, *Shank3* KO mice exhibited exacerbated synchronized grooming, potentially induced by specific ultrasound vocalizations from a member of the social group. Hence, chronic isolation may prevent this acquired behavior. Social touch has been shown to elicit aversive and atypical responses in the *Fmr1* KO mice, a model of Fragile X syndrome, and in a mouse model of maternal immune activation, potentially via the novel concept of brain endogenous noise[38–40]. If abnormal and aversive social touch processing also occurs in *Shank3* KO mice, it could elucidate the social and grooming phenotype induced by the life with several conspecifics.

Additionally, in mice, olfactory cues play a pivotal role in triggering social interaction, recognition, and motivation for conspecifics[30]. The overall olfactory preference remained unaffected, likely due to intact MOE function. However, fewer neurons in the VNO responded to urine and oxytocin, suggesting potential dysfunction in VNO-dependent social behaviors, such as maternal or male-pup aggression[30], and potentially impacting social interaction. While the implications of oxytocin and its receptor require further investigation, chronic isolation restored VSN responses in *Shank3* KO mice. Our results are consistent with the olfactory dysfunction in a rich odor environment observed in *Shank3* heterozygous mice[41]. Exploring the comprehensive sensory dysfunction in these mice, including aspects of social touch and olfaction, will enable us to understand their contribution to the ASD-like phenotype and the precise mechanism underlying the impact of chronic social isolation.

Two studies have reported positive effects of chronic social isolation in two mouse models of ASD. Chronic social isolation in *Nlgn3$^{R451C}$* KI males increased social interaction with females and normalized male aggression phenotype towards females, along with female urine exploration to WT levels[42]. The CX3CR1 receptor, implicated in ASD[11], exhibits increased protein levels due to social isolation in the prefrontal cortex, nucleus accumbens and hippocampus[43]. Chronic isolation reduced prepulse inhibition only in WT and increased traveled distance in both WT and *Cx3cr1* KO male mice[43]. Although not tested in males and females and on social interaction with the same sex, these studies align with our results in *Shank3* KO mice. However, further exploration in other mouse models of ASD is necessary to generalize the effects on ASD-like phenotype.

This unexpected result in *Shank3* KO mice suggests that, in the long term, individuals with ASD patients might be less impacted by social isolation during the pandemic than their neurotypical peers. Although our findings cannot be directly applied to autistic patients, this study suggests that vigilant monitoring of the social environment during the implementation of inclusive classrooms could enhance the success of integrating children with ASD. For example, ASD children might face challenges when exposed to an entire classroom of 20-30 children, and exposure to fewer children (of their choice) might facilitate the implementation and sustainability of inclusive classrooms. Here, we demonstrated that various social contexts influence WT social skills and *Shank3* KO is a suitable and useful model for testing behavioral interventions for ASD. Subsequent studies could investigate the most optimal number of cage mates (which may differ from the 4 WT animals demonstrated here) and the short- or long-term exposure to a mix of *Shank3* KO and WT on social interaction and stereotyped behavior.

**ACKNOWLEDGEMENTS**

Mouse breeding and care was performed by the trained technician of the rodent animal facility UE PAO (PAO, INRAE: Animal Physiology Facility, https://doi.org/10.15454/1.5573896321728955E12). We thank Dr Guillaume SALLE for the python script to analyze anymaze outputs for alternation patterns in the Y maze and Dr Flavie LANDOMIEL for her technical support.

This project has received funding from the European Research Council (ERC) under the European Union's Horizon 2020 research and innovation programme (grant agreement No. 851231). This work was supported by INRAE through the "SOCIALOME" project (PAF_29). LPP, AD and CG acknowledge the LabEx MabImprove (grant ANR-10-LABX-53-01) for their financial support of CG and AD PhD's co-fund.

**Data availability statement**

All the raw data are available in supplementary tables and the movies are available upon request.

**CONFLICT OF INTEREST**

The authors declare no financial conflict of interest.

**Author contributions**

**ABBREVIATIONS**

ASD autism spectrum disorder

KI knock-in

KO knockout

MOE main olfactory epithelium

OB olfactory bulb

VNO vomeronasal organ

VSN vomeronasal sensory neurons

WT wild-type

**FIGURE LEGENDS**

**Figure 1 WT mice exposed to a rich social environment display improved social skills**

In the Live Mouse Tracker, WT male and female mice raised in groups of 4 animals (dark gray; n = 48) displayed constant time in social approach (**A**), nose contacts (**B**), movements in contact (**C**) and isolated (**D**) across trials. In contrast, animals raised in group of 2 (light gray; n = 20) and 3 animals (middle gray; n = 33) spent less time in social approach (only group of 3), and nose contact (only group of 3), movements in contact and isolated across trials and compared to mice raised in group of 4 in the third trial. Data are presented as mean ± sd (**Table S2**). Robust linear model followed by pairwise comparisons using the estimated marginal means, with stars as trial effect and hash as housing effect (p = P adjusted in all tests). * or #: $p < 0.05$, ** or ##: $p < 0.01$, *** or ###: $p < 0.001$, **** or ####: $p < 0.0001$.

**Figure 2 chronic social isolation improves social interaction in WT mice**

(**A**) In the reciprocal test, WT group (gray) or WT isol (burgundy) animals spent similar time in nose contacts. In the three-chambered test, WT isol spent more in nose contacts with the mouse than WT group in the sociability phase (**B**), but not in the social novelty phase (**C**). Animals in both housing conditions preferred the mouse over the object or the novel mouse vs the familiar one. In the motor stereotypy test, housing conditions had no effect on the time spent in self-grooming (**D**) or number of head shakes (**E**). (**F**) WT isol traveled a larger distance than WT group. Data are presented as mean ± sd (**Table S3;** n = 34/housing condition). A-E; Kruskal-Wallis tests followed by Dunn post hoc tests, with stars as housing effect and hash as preference effect. * or #: $p < 0.05$, ** or ##: $p < 0.01$, *** or ###: $p < 0.001$. WT group, WT raised in groups; WT isol, WT exposed to chronic social isolation.

**Figure 3 Chronic social isolation increases oxytocin family gene expression only in the vomeronasal organ**

Relative quantitative expression of *Oxt*, *Oxtr* and *Avp* transcripts are 14 to 36 times increased in the VNO (**A**) of WT isol (burgundy; n = 6) compared to WT group (gray; n = 5), but not the housekeeping *Gapdh* transcripts or in MOE (**B**) or OB (**C**) tissues. Results are reported as mean ± standard deviation and a two-tailed Student's t test was conducted between groups. (**D**) Representative pictures of fluorescent *in situ* hybridization confirmed that oxytocin receptor transcripts (green puncta) were enriched in the VNO, both in the VNE and in the sustentacular cells of the NSE (arrowheads) of WT group (left) and WT isol (right) males; nuclei were counterstained with DAPI (blue). Scale Bar: 100µm. (**E**) Subcellular quantification of the FISH puncta using QuPATH. Each individual point represents quantification of one section, 5 sections per animal, n = 2 animals per condition, 7 sections of one animal for the control without probe. ONE-way ANOVA with *post-hoc* Tukey's T test. *: $p < 0.05$; **: $p < 0.01$; ****: $p<0.0001$. Ø, negative control with no probe; *Avp*, arginine vasopressin; MOE, main olfactory epithelium; NSE, non-sensory tissue; OB, olfactory bulb; *Oxt*, oxytocin, *Oxtr*, oxytocin receptor; VNE, vomeronasal sensory epithelium, VNO, vomeronasal organ; WT group, WT raised in groups; WT isol, WT exposed to chronic social isolation.

**Figure 4 Olfactory preference are not modified in WT mice exposed to chronic isolation or *Shank3* KO mice**

(**A**) In the olfactory preference test, the WT group (gray, n = 46) and isol (burgundy, n = 22) showed a preference for OT (1.5 µM) and urine over saline. *Shank3* KO mice (turquoise) displayed no significant urine preference (**B**) and no difference in latency (**C**) to sniff the paper with the opposite sex urine than WT (gray). The preference for OT was not significant in this cohort (n = 8 for OT and AVP 1.4 µM, n = 16 urine and paper). (**E**) Interestingly, the WT group showed a preference for urine of chronically isolated animals than urine from distinct animals raised in groups (n = 20). Data are presented as mean ± sd (**Table S4**). (**F**) The concentration of OT in the urine was not different in these two groups (isol, n = 5; group, n = 6). (**G**) Plasma levels of OT were increased in WT isol (n = 8) compared to WT group (n = 14). (**H**) As expected, OT concentration in the plasma and urine did not correlate (n = 7). Kruskal-Wallis tests followed by Dunn post hoc tests, with stars as housing effect. *: $p < 0.05$, **: $p < 0.01$, ****: $p < 0.0001$. AVP, arginine vasopressin; OT, oxytocin; Uisol, urine of mice exposed to chronic social isolation; WT group, WT raised in groups; WT isol, WT exposed to chronic social isolation.

**Figure 5 chronic social isolation restores social deficits and stereotyped behaviors in Shank3 KO mice**

(**A**) In the reciprocal test, WT isol (burgundy; n = 24-26) spent more time in nose contacts than WT group (gray; n = 47-62) and *Shank3* KO group (turquoise; n = 16) and KO isol (purple; n = 18) spent more time spent exploring a sex-, housing- and genotype matched conspecific. In the sociability (**B**) or the social novelty (**C**) phases of the three-chambered test, KO isol spent more time in nose contacts with the novel mice versus object or familiar mice than WT group and isol. All groups preferred the mouse over the object and the novel over the familiar mouse, except the *Shank3* KO group that did not show any preference for the novel mouse. In the motor stereotypy test, *Shank3* KO group showed increased time spent in self-grooming (**D**), number of head shakes (**E**) or time spent in synchronized grooming (**F**) than all the other conditions. The number of head shakes was not restored by isolation of the *Shank3* KO animals. A-F Data are presented as mean ± sd (**Table S5**). Kruskal-Wallis tests followed by Dunn post hoc tests. Calcium imaging revealed that an increased percentage of VSN cells of WT isol responded to sex-matched urine of animals raised in group (**G**), but not to oxytocin at 1 µM (**H**). The *Shank3* KO group displayed the opposite pattern with decreased percentage of cells responding to both stimuli, which was restored by chronic exposure to social isolation. (**I**) Representative calcium imaging traces of VSN responses to urine and oxytocin stimulations in the four groups. Scale bar: 1 min. G-H Fisher's exact test with n = 5-6 males per group. *: $p < 0.05$, **: $p < 0.01$, ***: $p < 0.001$, ****: $p < 0.0001$. KO group, *Shank3* KO raised in groups; KO isol, *Shank3* KO exposed to chronic social isolation; VSN, vomeronasal sensory neurons; WT group, WT raised in groups; WT isol, WT exposed to chronic social isolation.

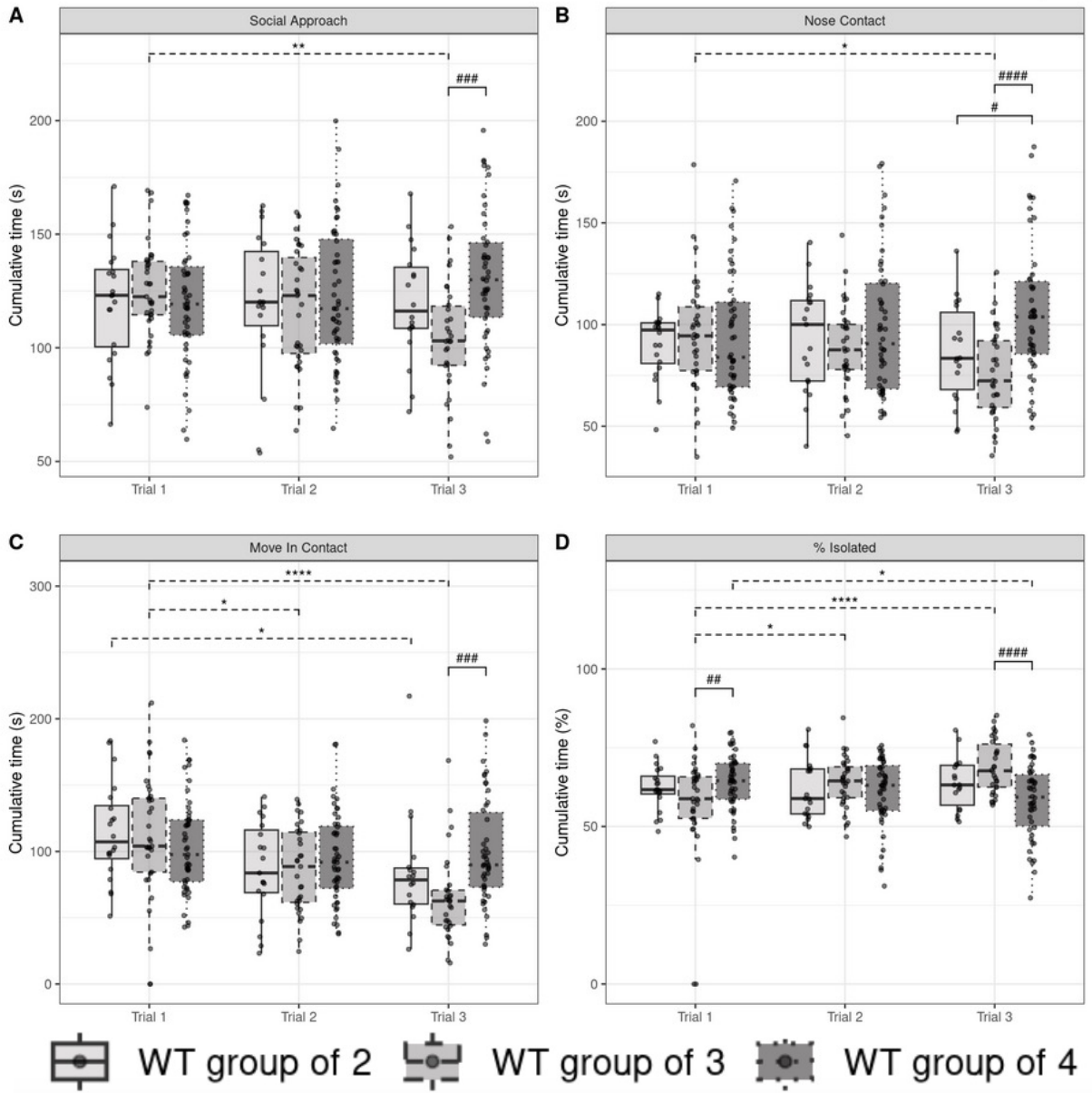

**Figure 1 WT mice exposed to a rich social environment display improved social skills**

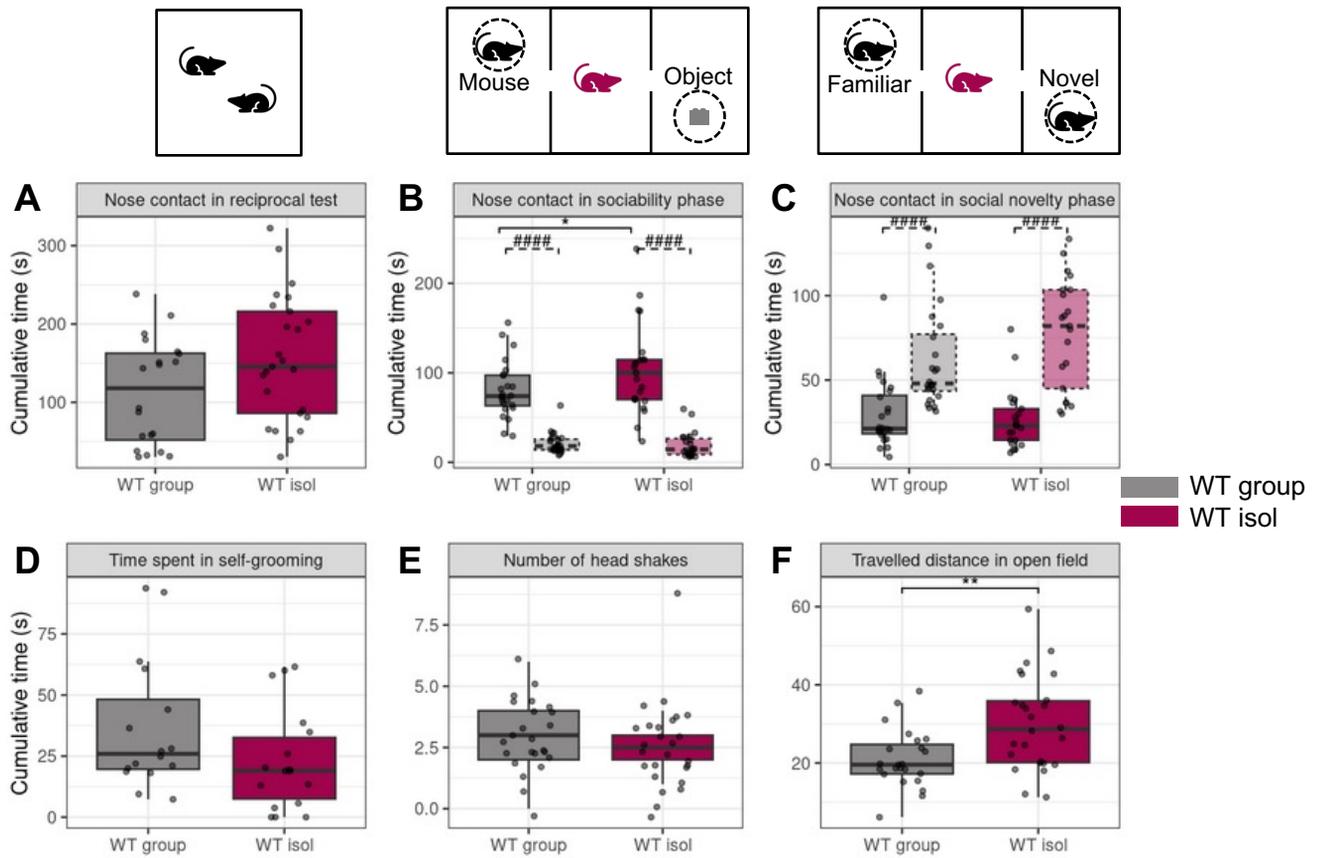

**Figure 2 chronic social isolation improves social interaction in WT mice**

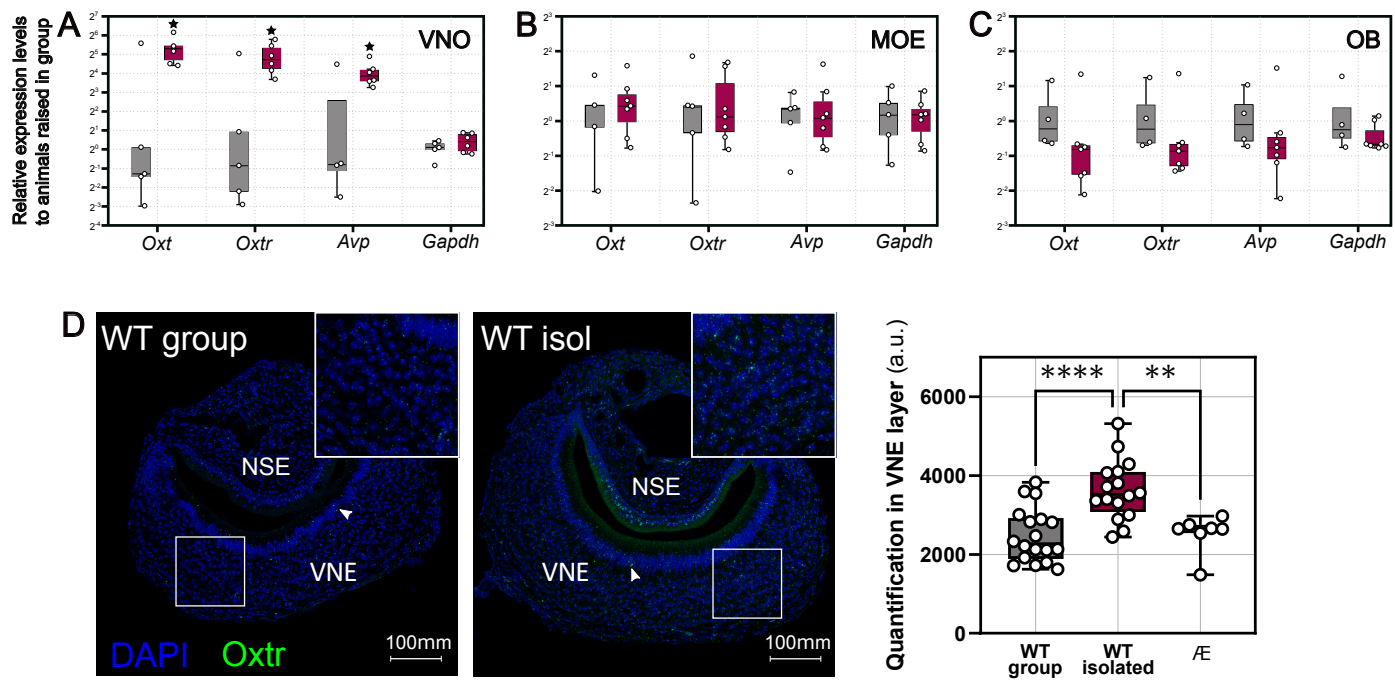

**Figure 3 Chronic social isolation increases oxytocin family gene expression only in the vomeronasal organ**

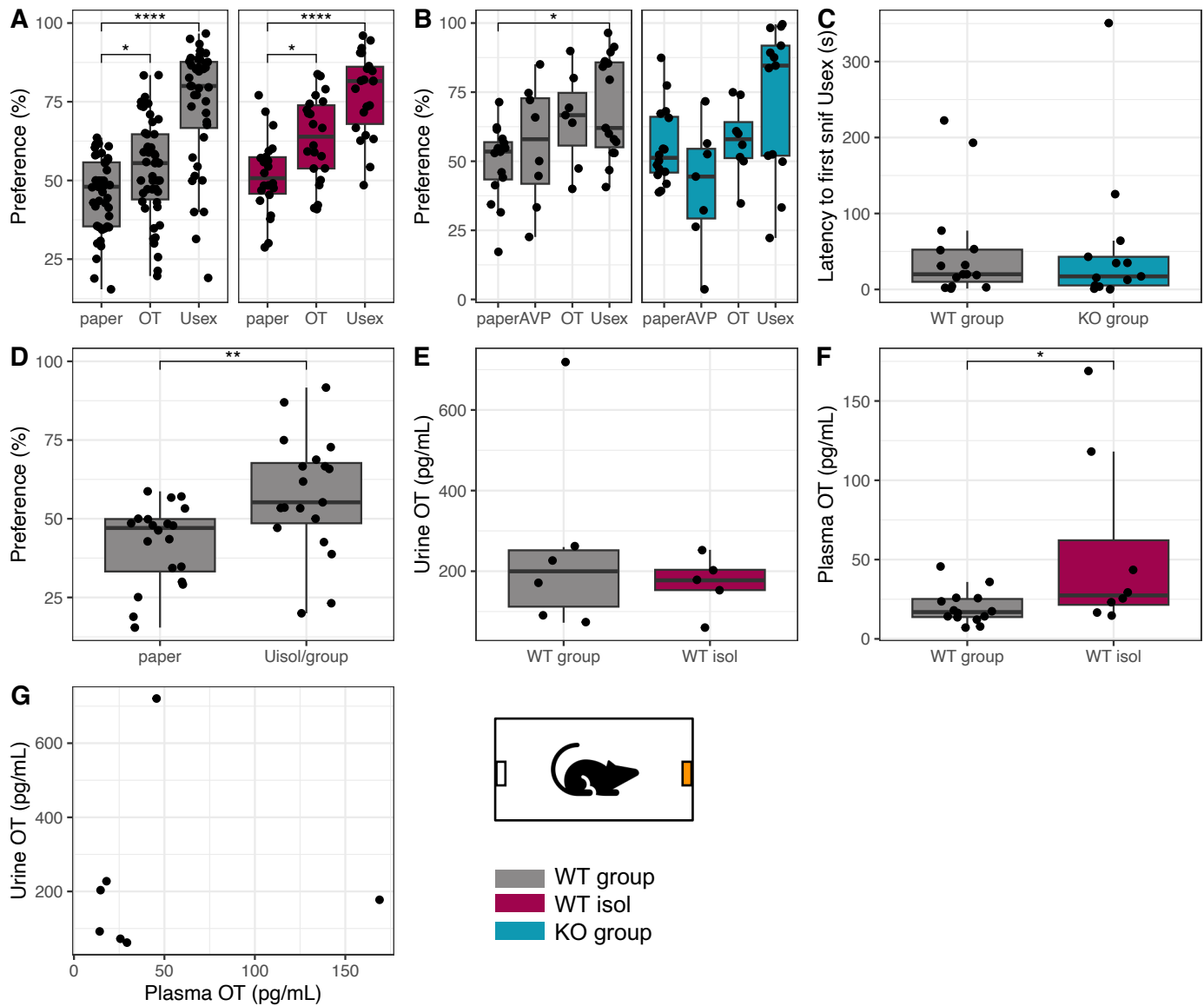

**Figure 4 Olfactory preference are not modified in WT mice exposed to chronic isolation or *Shank3* KO mice**

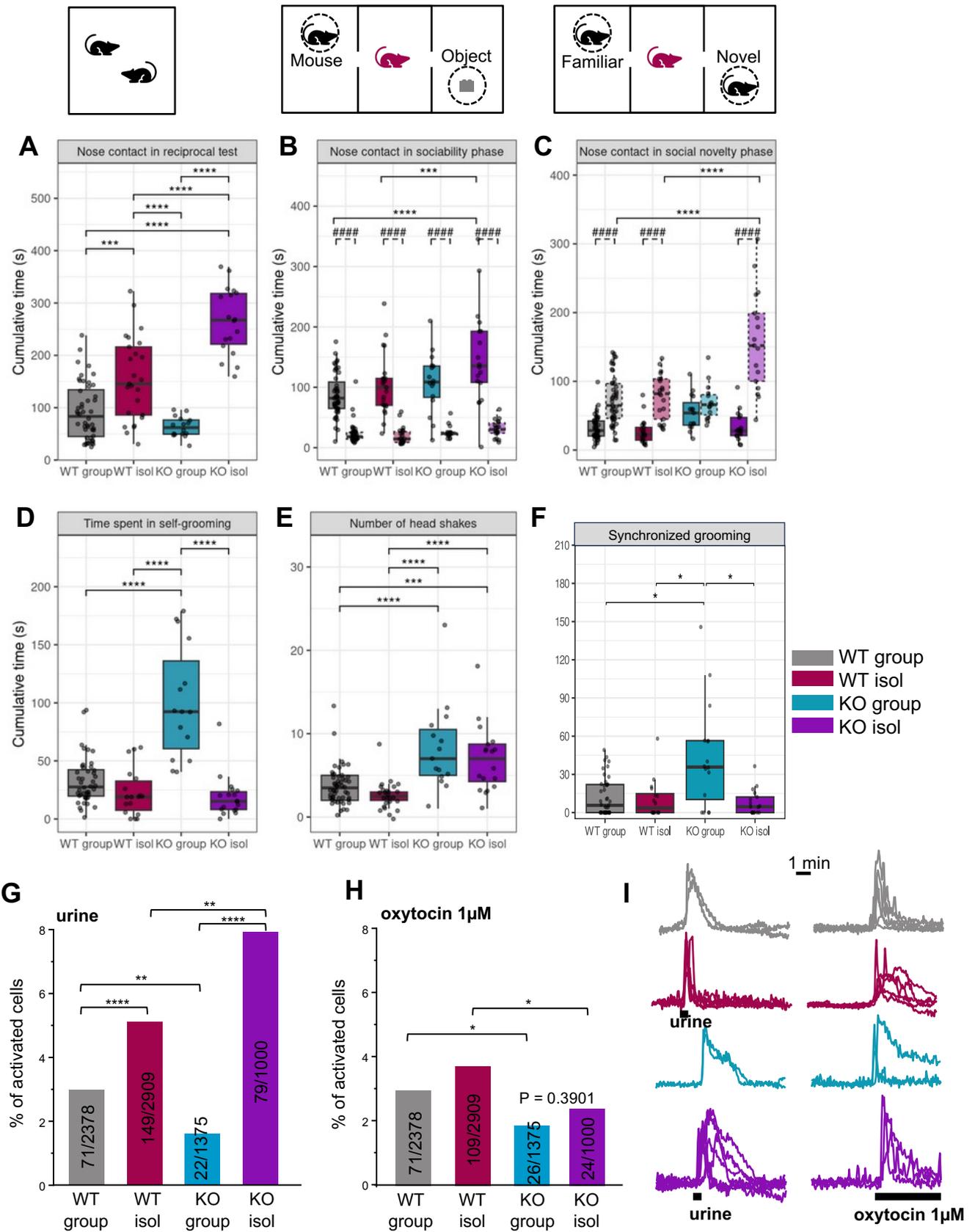

**Figure 5 chronic social isolation restores social deficits and stereotyped behaviors in *Shank3* KO mice**

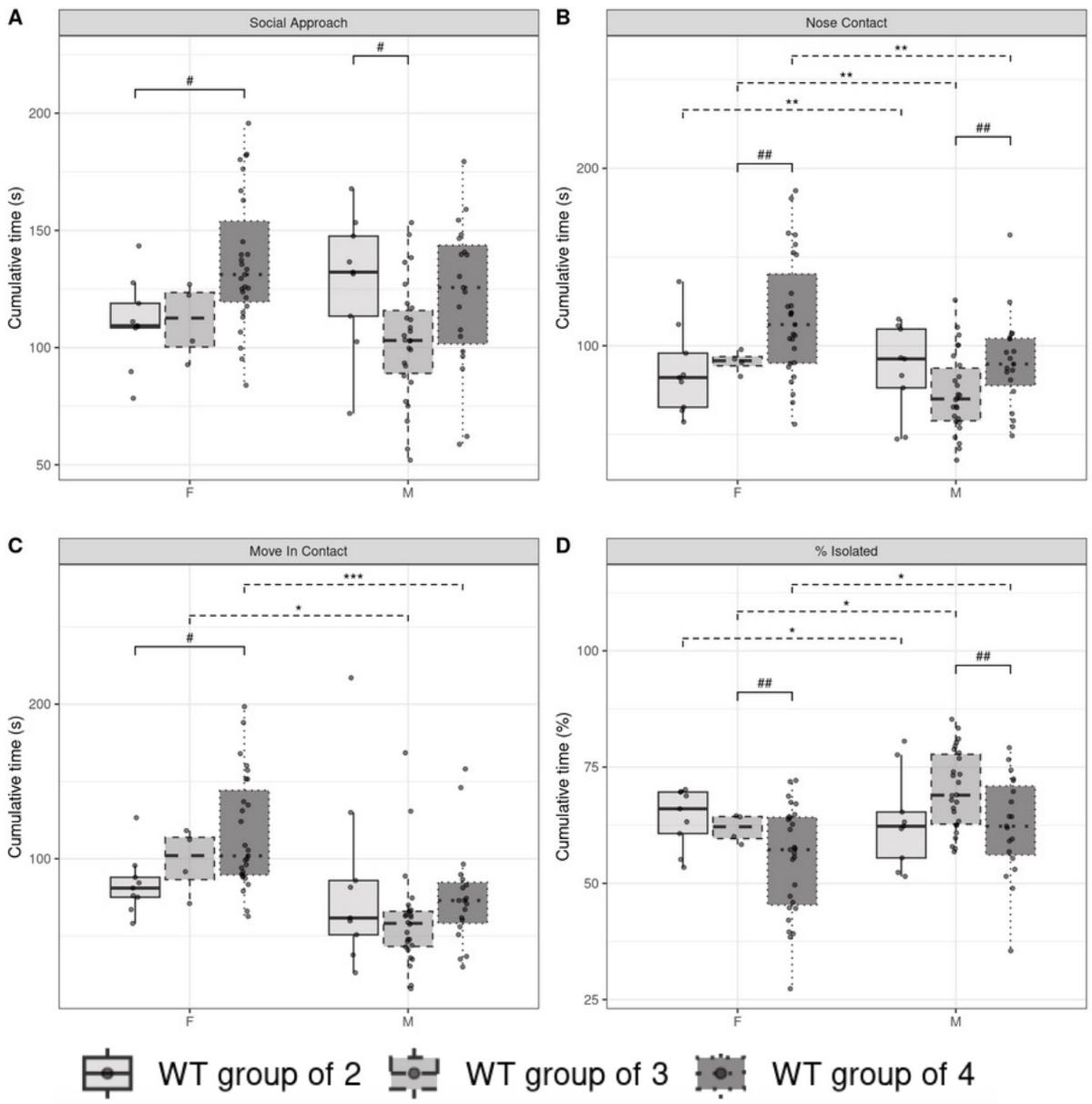

**Figure S1 Females exhibit overall enhanced social interaction than males**

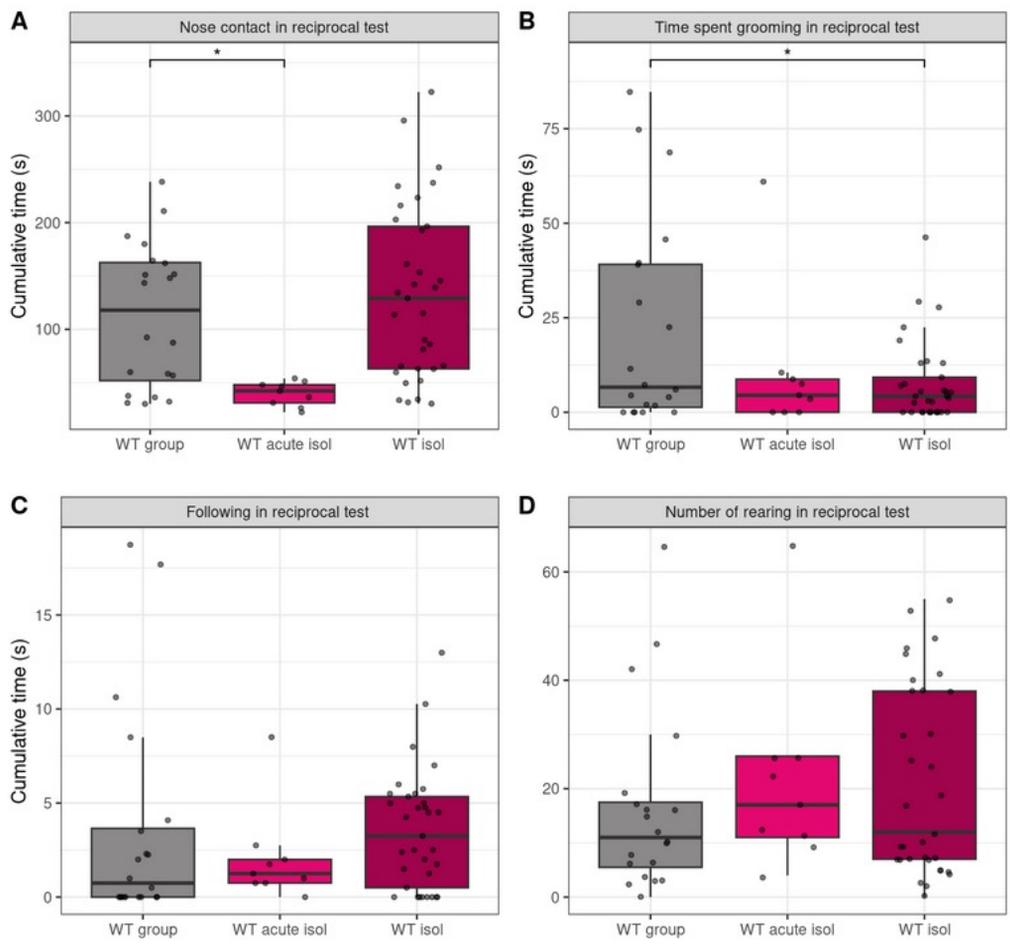

**Figure S2 acute social isolation induces social interaction deficits in WT mice**

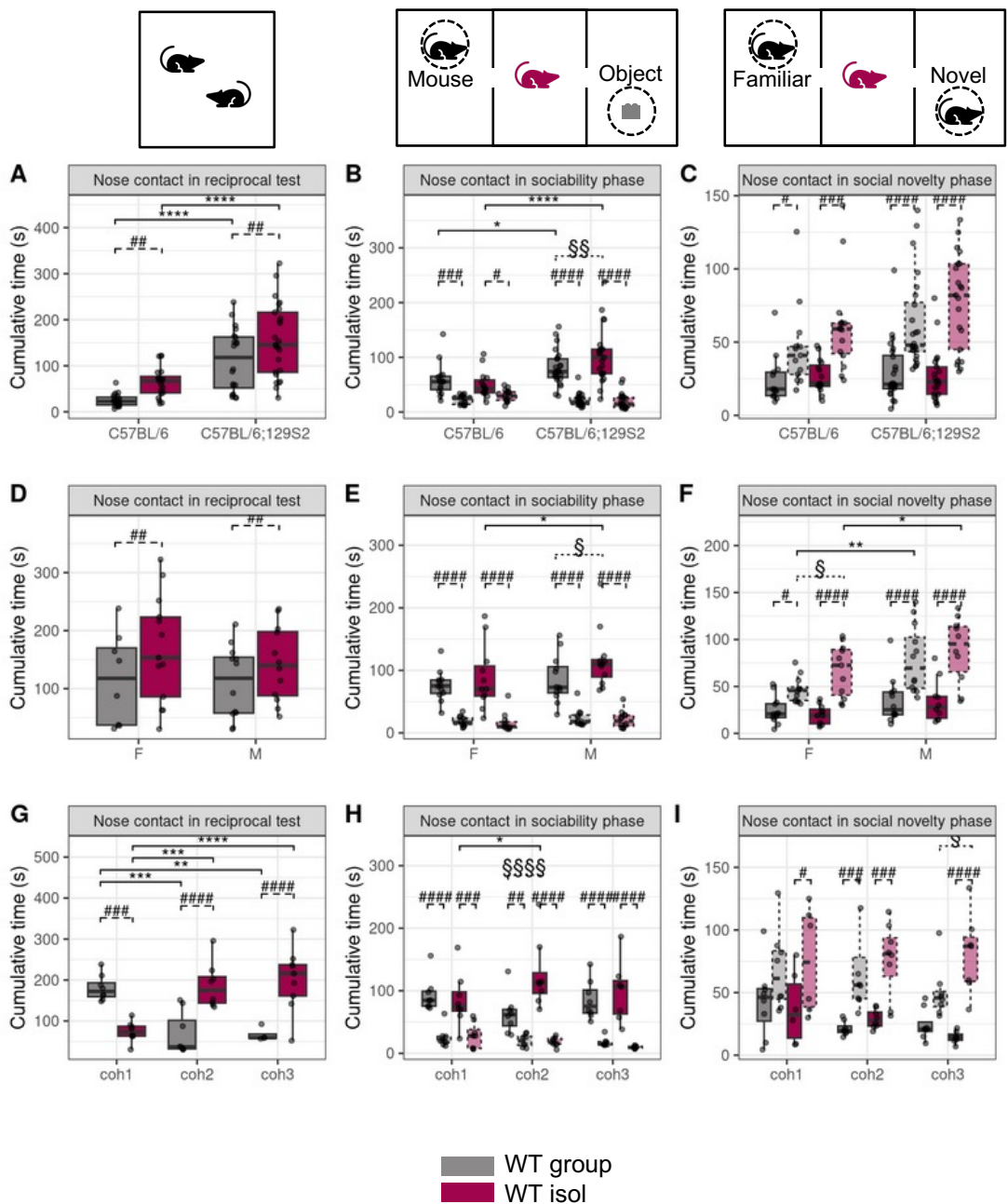

**Figure S3 chronic social isolation improves social interaction independently of background, sex and cohorts**

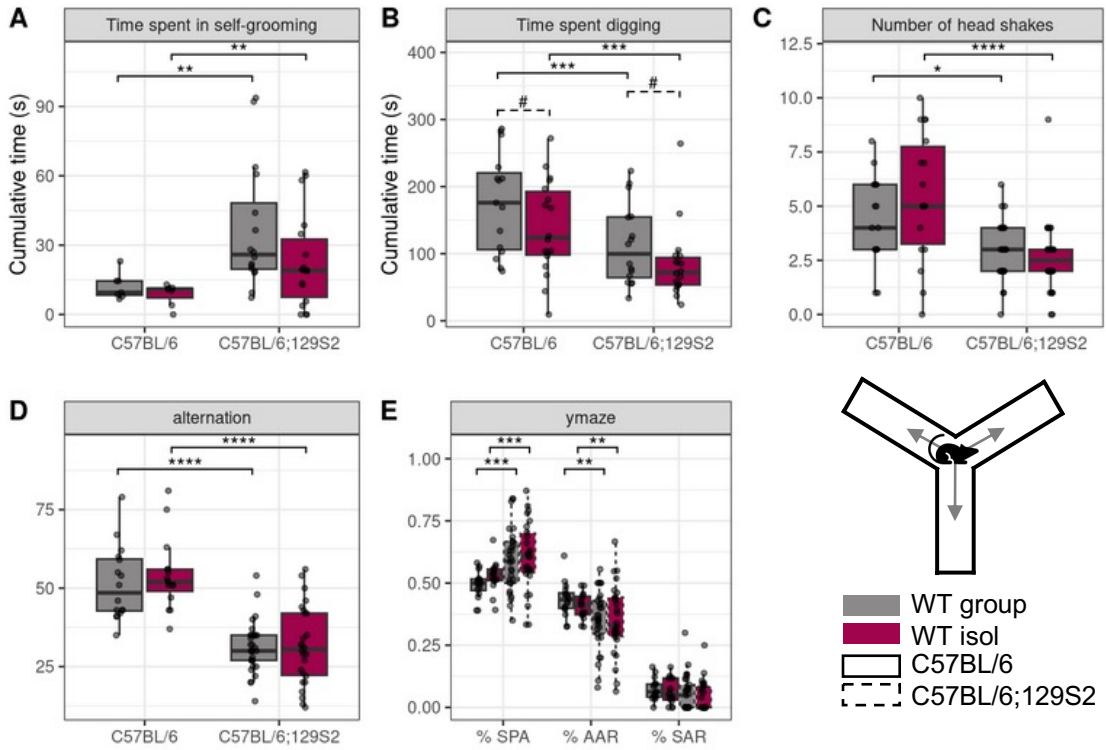

**Figure S4 Chronic social isolation does not induce stereotyped behaviors or cognitive impairments in WT mice**

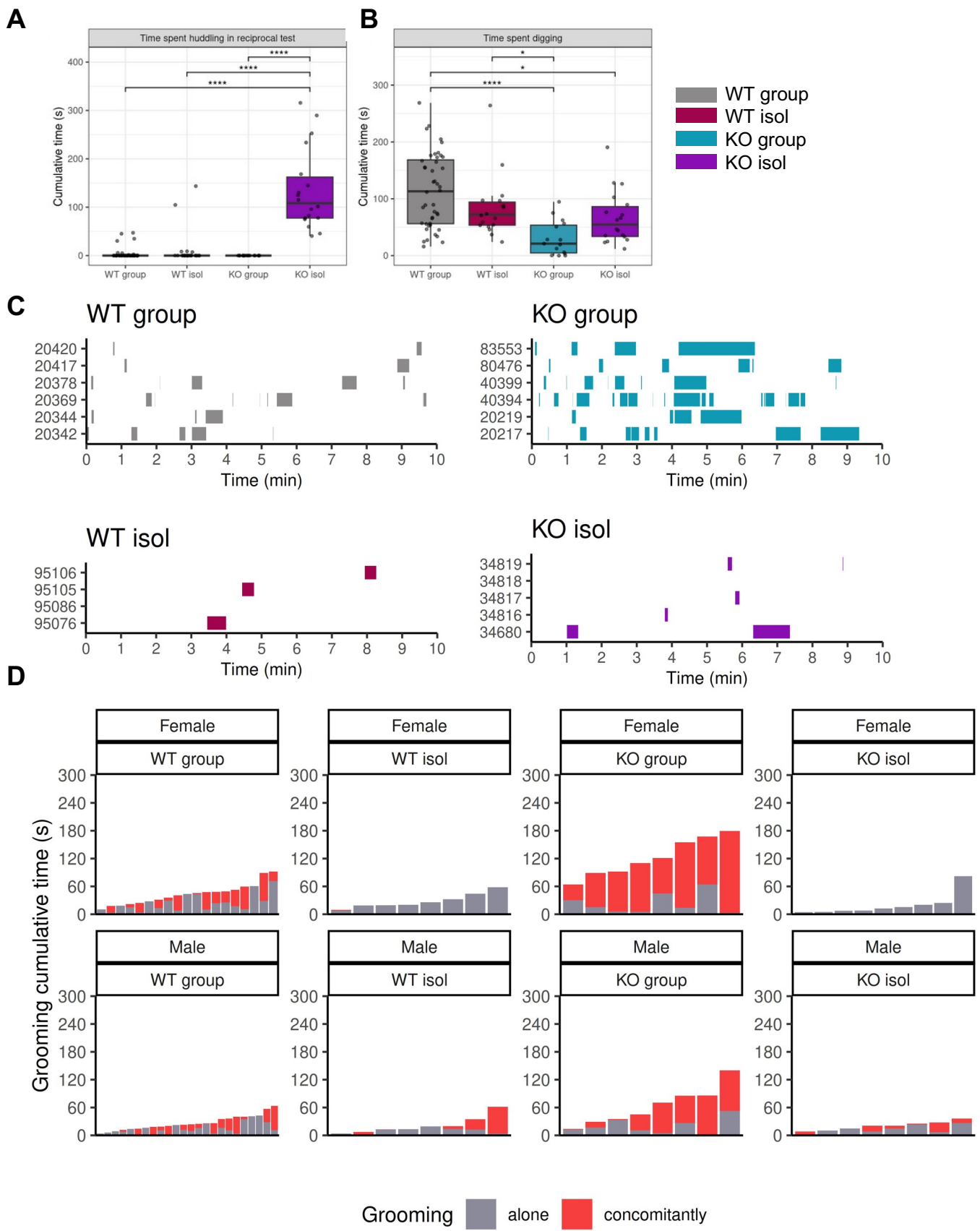

Figure S5 chronic social isolation normalizes stereotyped behaviors, including synchronized grooming in *Shank3* KO mice

# SUPPLEMENTARY INFORMATION

## SUPPLEMENTARY METHODS

### Behavior experiments

Timelines of the experimental design (O, object interaction; SI, reciprocal social interaction test, Y, spontaneous alternation in the Y maze; 3c, 3 chambered tests; ms, motor stereotypies; Olf, olfaction preference test)

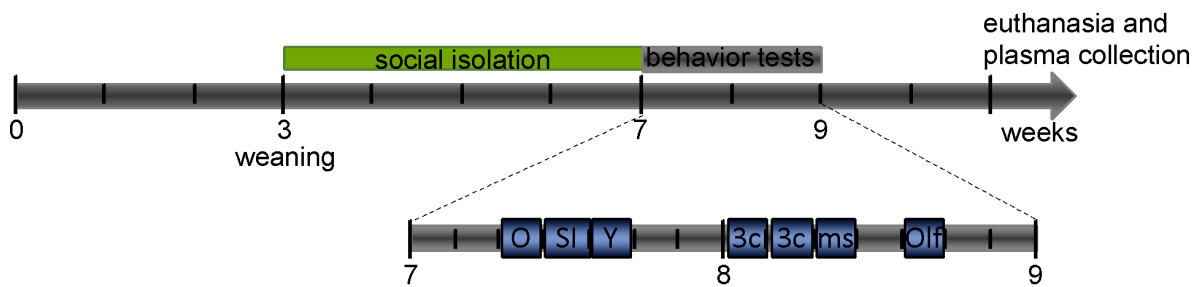

All mice (isolated or raised in groups) used in these experiments were raised in type 2 cage with the same material enrichment in a dedicated breeding room with controlled 21°C temperature and 50% humidity. In our facility, isolated mice were only exposed to deprivation of direct social olfactory and social touch from cage mate partners, but conserved olfactory, visual and auditory access to congeners located in the same room. Behavioral tests were carried out in a separated experimental behavioral room in low light intensity of 15 lux (40 lux for motor stereotypies) and standardized behavior apparatus[1,2], on mornings to avoid a circadian cycle effect and with one test per day, every 1-3 days. Resting days may vary between cohorts, although the same tests were performed on the same weeks. All mice were around 7-9 weeks of age from 100% C57BL/6J or mixed (exactly

50%/50%) C57BL/6J;129S2 backgrounds (Charles River, a supplier of Jackson Laboratory mouse strains). Following a heterozygous breeding scheme in the mixed C57BL/6J;129S2 background, a minimum of three WT and *Shank3* KO non-inbred couples generated the experimental cohorts.

All behavioral apparatus were built according to standard size and height in literature and manufacturer's instructions. Floors are composed of aluminum foil covered by a neutral gray and epoxy textured non-reflective gray floor to favor normal activity in the apparatus, well-being, optimal tracking and to reduce anxious-like behaviors.

All experiments were recorded using an USB black and white camera with 2.8-12 mm varifocal optic and ANY-maze tracking software (Stoelting, Ireland), Sony HD FDR-AX33 4K Camescopes or the Live Mouse Tracker system (LMT)[3]. All movies were automatically or manually scored using the ANY-maze tracking software and LMT script or Behavioral Observation Research Interactive Software[4], respectively. Animal exclusion criteria were either immobility (animals were immobile more than 30% of the total time of the test) or lack of exploration (animals did not visit at least twice the three different chambers, three arms, the center of open field arena or the two papers). ANY-maze software was configured to detect animal immobility using the 95% threshold, and animal entries in compartment or arms using the 80% of the animal body, for the three-chamber test and Y-maze.

### *Open field*

Two mice were placed for 10 minutes at 15 lux in an open field arena (46 x 46 cm; Ugo basile, Italy) with dark gray opaque partition walls with a Lego toy taped in the center of the arena (habituation for reciprocal social interaction test) using adhesive pasta. The cumulative time, the number of visits and mean duration spent in nose contact to the object, the traveled distance, the mean speed and the time spent in the squared central zone of the open field were automatically scored using the tracking ANY-maze software.

*Reciprocal social interaction test*

Two unknown sex-, genotype-, housing-matched mice were placed for 10 minutes at 15 lux in the same open field arena. The cumulative time, the number of visits and mean duration spent in nose contacts, self-grooming, self-grooming within 5 seconds following social contact, paw contacts, attacks, following, rearing, circling, huddling, vertical jumps and immobility were measured manually using BORIS, sex, genotype and housing conditions blinded to experimenter.

*Live Mouse Tracker*

Four unknown sex-matched WT mice (raised in 4 separate cages) on a mixed C57BL/6J;129S2 background were placed in the LMT (60 x 60 cm) arena with plexiglass transparent walls and recorded for 10 min at 15 lux. Each mouse was tested over 3 trials, with each time they encountered 3 other mice that they never met before or not for at least 7 days. According to previous study[5,6], social memory lasts between 1 and 14 days and therefore we considered that mice encountered only unknown mice in these three tests. A movie and an SQLite database file were generated for each run of 4 mice. The cumulative time in break contact, social approach, contact, get away, move in contact, move isolated, stretch attend posture (SAP), stop isolated and stop in contact parameters were scored automatically from SQLite database using Python scripts developed by de Chaumont et all[3]. 'Nose contacts' combine oral-oral and oral-genital contacts and 'following' combines follow zone, train of 2, 3 and 4 mice parameters. Rate periphery and rate isolated are the respective rate of the cumulative time in the periphery or spent isolated vs the total time of the run, respectively.

***The three-chamber social apparatus***

The three-chamber social apparatus (ugo basile, Italy) display three compartments of 20 x 40 x 22 (h) cm with 3 dark gray opaque and one transparent (masked by the other apparatus) PVC walls, separated by two transparent internal partitions, forming the opposite chambers A (up/left) and B (down/right), namely 'chA' and 'chB' and the central chamber 'chC' where the animal is initially placed at the start of each phase. Two mouse grid cages (7 cm diameter large, 15 cm tall, 7 mm spaced grid bars of 3 mm diameter and transparent lids) were placed at opposite locations within chA and chB chambers. The assay was divided in three phases of 10 minutes at 15 lux separated by 5 min intervals: the habituation phase with empty grid cages, the sociability phase with an unknown sex-matched WT interactor mouse versus a Lego "toy" and the social novelty phase with the sociability phase mouse versus a novel unknown WT mouse interactor replacing the toy. All interactor animals were trained at least twice for > 15 minutes on different days prior to the test to prevent any anxious-like bias. During the social phase, they were randomly assigned to the "B" (up/left) or "C" (down/right) grid cages to prevent any spatial bias in the behavior room. For the habituation phase, the cumulative time, the number of entries/visits and mean duration spent in close contact for B or C empty grid cage zones, in chA, chB or chC chambers, immobility and the total traveled distance were scored automatically using the tracking ANY-maze software. For the social and social novelty phases, the cumulative time, the number of entries/visits and mean duration spent in nose contacts for interactor or close contact grid cages containing the toy, self-grooming, in chA, chB or chC chambers, and immobility were measured manually, as anymaze software did not score accurately nose contacts. To note, when an experimental animal explored the top of a grid cage, the time was not considered as a close interaction unless the interactor animal was also located at the top of the grid cage. An entry in a chA or chB is validated only when the mouse enters its 4 paws.

***Spontaneous alternation pattern in the Y-maze***

Spontaneous alternation pattern was used to measure cognitive flexibility and perseverative behavior in the Y-maze spatial working memory task[7,8]. Each Y-maze (ugo basile, Italy) displays three standardized dimensions of PVC gray opaque arms (l 35 x w 5 x h 10 cm) that can be disassembled and closed with the included doors. Each mouse was placed for 5 minutes at 15 lux randomly in one of the arms to prevent any spatial bias in the behavior room. The three arms were randomly identified as "A", "B", and "C". The percentage of spontaneous alternation (SPA), alternative arm returns (AAR) and same arm returns (SAR) determined from a triplet of arm entries was automatically scored using the automatic ANY-maze software, as well as the total traveled distance, the mean speed of the experimental mouse, the number of entries in the arms, the cumulative time and the mean duration of the time spent in each arm. An entry in an arm was validated only when the mouse entered with its 4 paws.

***Motor stereotypies***

Stereotyped and repetitive behaviors[1,2] were recorded in a type I cage filled with 4-5 cm deep layer of fresh sawdust litter and covered with a steel grid for 10 minutes at 40 lux. Six to eight experimental mice were placed next to each other during each recording. The cumulative time, the number and mean duration of repetitive behaviors (self-grooming and digging) and motor stereotypies (vertical/horizontal jumps, circling, head shakes and scratching episodes) were manually scored as well as the cumulative time spent immobile and the number of rearing as a measure of spatial exploration.

*Olfactory preference test*

Mice were placed in type I cage with a thin layer of fresh sawdust litter and steel grid filled, containing two squared (1.5 x 1.5 cm whatman) papers stuck on the walls with double face tape (5 cm from the bottom) at the two extremities of the transparent walls. Mice were allowed to explore this new environment for habituation, 10 min at 15 lux. After habituation, no more than two consecutive trials of 10min were performed per day with papers randomly soaked with 20µL of oxytocin (OT; Tocris bioscience reference 1910; 30 µg final per mouse or around 1.5mM) or vasopressin (AVP; Tocris bioscience reference 2935; 30 µg final per mouse or around 1.4mM) *vs* saline (SAL; NaCl 0,9 %), urine of sex-matched isolated animals (Uisol) *vs* urine of sex-matched animals raised in groups (Ugroup) or urine of opposite sex congener (Usex) *vs* SAL. Compounds were diluted in SAL for 100% C57BL/6J background whereas they were diluted in a (50-50) mix of saline and urine of unknown and sex matched congener for the mixed C57BL/6J;129S2 background. Urines of isolated or grouped animals were collected either during habituation phases of 3-chambered, open field or empty grid cages from different animals than the experimental animals used for olfaction preference tests. Urine samples of males and females from the different housing conditions were pooled separately and kept at -20°C until further use. To note, the time mice spent playing or biting the papers was considered as time spent in olfaction. Olfaction preference was calculated as the percentage of time spent sniffing and/or approaching (< 0.5 cm) the test paper over the 10 min period versus the total time exploring the test and reference papers. In order to prevent mouse side preference in the behavior room, reference paper in the habituation phase was randomly assigned and papers were soaked with compound or urine on this reference paper. Olfaction preference was calculated according to this reference paper.

*Synchronized grooming*

Although similar results were observed in the olfactory preference test, the synchronized grooming behaviors of mice were evaluated in the motor stereotypy assay. Actually, four to six mice were recorded simultaneously in the same room using two to three cameras, respectively. The scoring of the behaviors was performed using BORIS software. To ensure the synchronism of the scoring, potential delays between movies of the same run were taken into account. The start of the scoring of a run was defined as the time of the first rearing of a mouse. The stop of the scoring was set ten minutes after this start. Cumulative time of a behavior was computed as the sum of all the behavior's event durations. Synchronized grooming was measured as the cumulative time spent in self-grooming concomitantly with another experimental mouse for each run. Comparison of the cumulative times spent in digging and in grooming between all the different groups was performed using Wilcoxon signed-rank tests with Bonferroni corrections. Data management, graphical evaluation and statistics were performed using R (R version 4.3.1 (2023-06-16), Vienna, Austria).

*Plasma and urine oxytocin concentration*

Plasma or urine oxytocin concentration was determined by EIA (kit ADI-901-153, Enzo Life Sciences) after solid-phase extraction using Sep-Pack C18 cartridges (Waters). 120µL of plasma or 150µL of urine were extracted then samples were reconstituted in 120µL or 150µL assay buffer. Measurements were performed in duplicate. The assay sensitivity was 7.8pg/ml.

For urine samples, creatinine concentrations were also measured in each sample (kit : 80350, Crystal Chem) and oxytocin values were expressed as pg/mg of creatinine.

*Quantitative PCR*

Total RNAs were extracted according to the manufacturer instructions (Zymo Research Corporation, kit Direct-zol RNA Microprep) and quantified using a nanodrop (Thermo Fisher Scientific, Waltham, MA). MOEs, VNOs and OBs cDNAs were generated respectively from 450 ng, 320 ng and 115 ng of total RNAs using the SuperScript III kit (Invitrogen) according to the manufacturer protocol. MOEs and VNOs qPCRs were performed using a 1/50 cDNA dilution while a 1/25 dilution was applied to OBs. Primers used were validated for the corresponding genes (Table S1). All assays were performed on a CFX384™ Real Time PCR Detection System. Each PCR reaction contained 6.25 µl of the 2X ONEGreen Fast qPCR premix (Ozyme, USA), 1 µmol/l of each primer, 1 µl of diluted template DNA and water to a total reaction volume of 12.5 µl. Reactions were carried out with an initial incubation at 95 °C for 3 minutes, followed by 40 cycles of denaturation at 95 °C for 5 seconds, annealing at 60 °C for 15 seconds, and elongation at 60 °C for 30 seconds. All reactions were carried out in triplicate and negative controls, containing water instead of template DNA, were included in every PCR run. A melting curve analysis was carried out to ensure the specificity of the amplification products. The amplification efficiencies of all qPCR assays ranged from 91% to 100.7%.

Analysis was performed with the BioRad CFX Maestro 2.3 software. Normalized relative expression calculated using control samples and reference targets (ΔΔCq) was the gold standard method for this study.

*RNAscope In situ hybridization*

Staining for Oxtr mRNA was performed using multiplex fluorescent in situ hybridization (ISH). Mice were anesthetized with 100 mg/kg ketamine and 5 mg/kg xylazine and perfused transcardially with 0.9% saline solution followed by 4% paraformaldehyde (PFA) in 0.1 M phosphate buffer saline (PBS). VNO tissue was dissected, postfixed overnight in 4% PFA and later cryoprotected in PBS containing 30% sucrose. Samples were embedded in Tissue-Tek O.C.T. compound, snap-frozen in cold isopentane and processed on a Leica CM 3050S cryostat. VNOs were cut in 16-µm thick coronal sections and were directly mounted on SuperFrost Plus slides glasses (Thermo Scientific). RNAscope Fluorescent Multiplex V2 labeling kit (ACDBio 323110) using mm-Oxtr probe (ACDBio xxx) was optimized from the manufacturer's protocol. Briefly, sections were allowed to dry, were post-fixed in 4% PFA for 15 min at room temperature, treated with 3% H2O2 in 1X PBS for 15 min, submitted to antigen retrieval in 1X antigen retrieval buffer at 95°C for 5 minutes and rinsed in water and PBS. Slides were then incubated in ethanol 70% and ethanol 100% and allowed to dry for 5 min. In situ hybridization steps: the slides were then incubated with Protease III for 30 min at 40°C, covered in parafilm. After that they were washed 2 times 5 min each in 1X wash buffer, and incubated with the probes (diluted 1:50 on Probe Diluent Solution) for 2 h at 40°C. Slides were washed 2 times 5 min each in 1X wash buffer and incubated with the Amplification solutions 1 2 and 3 according to the manufacturer's protocol, but covered with parafilm at each incubation step. For the signal development, the slides were incubated with HRP RNAscope reagent for 15 min at 40°C, washed 2 times, incubated with TSA Plus Biotin (Akoya) 1:50 (In Amplification Buffer, Akoya), then washed 2 times, then incubated for 15 min at 40°C with Cy2-labeled streptavidin 1:500 (Jackson ImmunoResearch). Slides were then subjected to HRP blocking with the HRP blocking reagent for 15 min at 40°C, washed and either counterstained with Hoechst 33258 and mounted with antifade fluorescent mounting medium (Dako). Fluorescent images were captured using sequential laser scanning confocal microscopy (Zeiss LSM-780). Details of quantification

*Calcium imaging*

Ca2+ imaging of freshly dissociated VSNs was performed as previously described[9,10]. VNO epithelium was detached from the cartilage and minced in PBS at 4°C. The tissue was incubated (20 min at 37°C) in PBS supplemented with papain (0.22 U/ml; Worthington) and DNase I (10 U/ml; Thermofisher), gently extruded in DMEM (Invitrogen) supplemented with 10% FBS, and centrifuged at 100 × g (5 min). Dissociated cells were plated on coverslips previously coated with concanavalin-A type V (0.5 mg/ml, overnight at 4 °C; Sigma). Cells were used immediately for imaging after loading them with fura-2/AM (5 µM; Invitrogen) for 30 min. Cell-containing coverslips were placed in a laminar-flow chamber (Warner Instruments) and constantly perfused at 22 °C with extracellular solution Hank's balanced salt solution (HBSS, Invitrogen) supplemented with 10 mM Hepes (2-[4-(2-hydroxyethyl)piperazin-1-yl]ethanesulfonic acid). Cells were alternately illuminated at 340 and 380 nm, and light emitted above 510 nm was recorded using a C10600-10B Hamamatsu camera installed on an Olympus IX71 microscope. Images were acquired at 0.25

Hz and analyzed using ImageJ (NIH), including background subtraction, region of interest (ROI) detection and signal analyses. ROIs were selected manually and always included the whole cell body. Peak signals were calculated from the temporal profiles of image ratio/fluorescent values. Results are based on recordings from 5 - 6 mice for each condition and genotype. Cells were stimulated successively and in random order using bath application with control HBSS-Hepes buffer, 1 µM oxytocin (Sigma), and urine mixed from at least three different C57Bl/6 adult male mice and diluted to 1:100. To identify and mark Ca2+ responses to the applied stimuli, we followed these criteria: (1) A response was defined as a stimulus-dependent deviation of fluorescence ratio that exceeded twice the SD of the mean of the baseline fluorescence noise. (2) Cells showing a response to the control buffer were excluded from analysis. (3) A response had to occur within 1 min after stimulus application. In time series experiments, ligand application was repeated to confirm the repeatability of a given Ca2+ response.

**SUPPLEMENTARY LEGENDS**

**Figure S1 Females exhibit overall enhanced social interaction than males**

In the third trial of the Live Mouse Tracker, WT females raised in groups of 4 animals (dark gray; n = 28) displayed increased time in social approach (**A**), nose contacts (**B**), movements in contact (**C**) and spent less time isolated (**D**) compared to females raised in groups of 2 (light gray; n = 10) and 3 (middle gray; n = 6). Males raised in groups of 4 (n = 20) showed enhanced time in nose contact and less time isolated than males raised in groups of 3 (n = 27), but not with 2 (n = 10). Overall, females showed increased time present in nose contacts, in movements in contact and isolated than males. Data are presented as mean ± sd (Table S2). Robust linear model followed by pairwise comparisons using the estimated marginal means, with stars as sex effect and hash as housing effect (p = P adjusted in all tests). * or #: $p < 0.05$, ** or ##: $p < 0.01$, *** or ###: $p < 0.001$, **** or ####: $p < 0.0001$.

**Figure S2 acute social isolation induces social interaction deficits in WT mice**

(**A**) In the reciprocal test, WT acute isol males and females (pink, n = 9) spent less time in nose contact than WT group (gray, n = 34) and isol (burgundy, n = 34). (**B**) WT isol spent less time in self-grooming and none of the conditions was different in time present in following (**C**) or in the number of rearing episodes (**D**). Data are presented as mean ± sd (Table S3). Kruskal-Wallis tests followed by Dunn post hoc tests, with stars as housing effect. *: $p < 0.05$. WT group, WT raised in groups; WT isol, WT exposed to acute (2 hours) social isolation; WT isol, WT exposed to chronic social isolation.

**Figure S3 chronic social isolation improves social interaction independently of background, sex and cohorts**

(**A**, **D**, **G**) In the reciprocal test, WT isol (burgundy) males and females (**D-F**), in the C57BL/6 (**A-C**) or C57BL/6;129S2 backgrounds (**A-I**) and across different cohorts, except cohort 1 (**G-I**), spent overall more time in nose contacts than WT group (gray). C57BL/6;129S2 background showed enhanced social skills in this test than C57BL/6 background for both housing conditions. (**B**, **E**, **H**) In the sociability phase of the three-chambered test, WT isol animals spent more time in contact with the mouse over the object, which was found especially in males and cohort 2. Again, C57BL/6;129S2 background exhibited increased nose contacts compared to C57BL/6 background for both housing conditions. In all conditions, animals showed preference for the mouse over the object. (**C**, **F**, **I**) In the social novelty phase, only WT isol females spent more time in contact with the novel mouse over the familiar one, whereas the preference was intact in almost all conditions. Males displayed enhanced interaction compared to females. Data are presented as mean ± sd (Table S3). Kruskal-Wallis tests followed by Dunn post hoc tests, with stars as background, sex or cohort effect, hash as housing condition effect and § as preference effect. *, § or #: $p < 0.05$, **, §§ or ##: $p < 0.01$, ***,§§§ or ###: $p < 0.001$, ****,§§§§ or ####: $p < 0.0001$. coh, cohort; F, females; M, males; WT group, WT raised in groups; WT isol, WT exposed to chronic social isolation.

**Figure S4 Chronic social isolation does not induce stereotyped behaviors or cognitive impairments in WT mice**

In the motor stereotypy test, WT isol exhibited no difference in the time spent in self-grooming (**A**), reduced time spent digging (**B**) and no difference in the number of head shakes (**C**). C57BL/6;129S2 background spent more time in self-grooming, less time digging and had reduced number of head shakes compared to C57BL/6 background. In the Y maze, housing had no effect on the number (**D**) or pattern (**E**) of alternations. C57BL/6;129S2 displayed reduced number of alternations and showed enhanced percentage of SPA, concomitant with reduced AAR compared to C57BL/6 background. Data are presented as mean ± sd ([Table S3](#)). Kruskal-Wallis tests followed by Dunn post hoc tests, with stars as background effect and hash as housing effect. * or #: p < 0.05, **: p < 0.01, ***: p < 0.001. AAR, alternate arm returns; SAR, same arm returns; SPA, spontaneous alternation; WT group, WT raised in groups; WT isol, WT exposed to chronic social isolation.

**Figure S5 chronic social isolation normalizes stereotyped behaviors, including synchronized grooming in Shank3 KO mice**

(**A**) In the reciprocal test, KO isol (purple; n = 18) spent a lot of time time spent huddling compared to WT isol (burgundy; n = 24-26), WT group (gray; n = 47-62) and *Shank3* KO group (turquoise; n = 16). In the motor stereotypy test, *Shank3* KO group showed reduced time spent digging (**B**), which was not restored by chronic isolation. (**C**) Representative Gantt charts highlighted the pattern of synchronized self-grooming in the KO group. (**D**) Individual bar plots illustrated that males and females in the KO group spent more time in concomitant grooming than grooming alone and in other conditions. Data are presented as mean ± sd ([Table S5](#)). Kruskal-Wallis tests followed by Dunn post hoc tests. *: p < 0.05, ***: p < 0.001, ****: p < 0.0001. KO group, *Shank3* KO raised in groups; KO isol, *Shank3* KO exposed to chronic social isolation; WT group, WT raised in groups; WT isol, WT exposed to chronic social isolation.

**Table S1 list of validated primers**

**Table S2 raw, mean, standard deviation and statistics of social interaction in the Live Mouse tracker of WT mice**

**Table S3 raw, mean, standard deviation and statistics of behaviors in WT exposed to acute and chronic social isolation**

**Table S4 raw, mean, standard deviation and statistics of olfactory assays in WT and *Shank3* KO mice**

**Table S5 raw, mean, standard deviation and statistics of behaviors in WT and *Shank3* KO mice exposed to chronic social isolation**

**SUPPLEMENTARY REFERENCES**

1 Crawley JN. Behavioral phenotyping strategies for mutant mice. *Neuron* 2008; **57**: 809–818.

2 Crawley JN. Translational animal models of autism and neurodevelopmental disorders. *Dialogues Clin Neurosci* 2012; **14**: 293–305.